\documentclass[11pt,a4paper]{article}

\usepackage[margin=1in]{geometry}
\usepackage{times}
\usepackage{amsmath}
\usepackage{amssymb}
\usepackage{graphicx}
\usepackage{booktabs}
\usepackage{hyperref}
\usepackage{url}
\usepackage{cite}
\usepackage{float}
\usepackage{array}
\usepackage{lineno}
\usepackage{setspace}
\usepackage{titlesec}
\usepackage{abstract}
\usepackage{xurl}
\usepackage{hyperref}
\usepackage{amsmath}
\usepackage{amssymb}
\usepackage{amsthm}

\usepackage{graphicx}
\usepackage{booktabs}
\usepackage{multirow}
\usepackage{xcolor}
\usepackage{algorithm}
\usepackage{algpseudocode}
\hypersetup{
    colorlinks=true,
    linkcolor=blue,
    citecolor=blue,
    urlcolor=blue
}

\titlespacing*{\section}{0pt}{12pt}{6pt}
\titlespacing*{\subsection}{0pt}{10pt}{4pt}

\title{
    \Large\textbf{
   Autonomous OSS Threat Detection via Taxonomy-Aligned LLMs
    }
}

\author{
\begin{tabular}{c}
\textbf{Md. Robiul Islam Niloy}\\
Department of Computer Science and Engineering\\
BRAC University, Dhaka, Bangladesh\\
\texttt{md.niloy26643@gmail.com}
\end{tabular}
}

\date{October 2026}

\begin{document}

\maketitle

\begin{abstract}
Open source software (OSS) ecosystems face escalating threats 
from sophisticated supply chain attacks including typosquatting, 
dependency confusion, Trojan Source obfuscation, malicious build 
injection, and CI/CD pipeline poisoning. Sonatype documented 
over 1.2 million cumulative malicious packages across major 
registries as of 2026, yet existing detection approaches rely 
on signature-based tools and rule-based systems that lack the 
semantic understanding needed to generalize across attack 
variants and emerging threat patterns. In this paper we propose 
a taxonomy-aligned large language model framework for automated 
detection and multi-class classification of OSS supply chain 
threats.

We introduce a structured AV-xxx threat taxonomy covering five 
semantically distinct attack categories and construct a curated 
dataset of 999 verified real-world OSS supply chain incidents 
sourced from GitHub Security Advisories, CISA alerts, 
government security databases, and commercial threat 
intelligence reports spanning 2018 to 2026 across eight 
package ecosystems. Using taxonomy-aligned prompt engineering 
with GPT-4, our framework achieves 97.0\% multi-class 
classification accuracy and 97.0\% macro F1 score across 
all five threat categories.

Comprehensive evaluation against eight methods --- including 
five traditional machine learning baselines, one zero-shot 
open source LLM, and two fine-tuned neural models --- reveals 
a surprising finding: fine-tuned Llama 3.1 8B (70.5\%) and 
SecRoBERTa (77.5\%) both underperform simple TF-IDF 
classifiers (82.3\%), while Mistral 7B without taxonomy 
alignment achieves only 65.7\%. These results confirm that 
taxonomy-aligned prompting rather than model scale, domain 
pretraining, or fine-tuning is the critical performance 
factor. Cross-ecosystem analysis further reveals that 
classification difficulty varies substantially across 
ecosystems, with npm achieving only 65.5\% baseline 
accuracy compared to 100.0\% on smaller ecosystems, 
and that GitHub Actions incidents exhibit a critical 
accuracy-F1 divergence (82.4\% accuracy, 0.367 macro F1) 
indicating systematic class imbalance in baseline 
predictions --- precisely the ecosystems where LLM-based 
contextual reasoning provides the greatest benefit.

Our dataset, fine-tuning scripts, and evaluation code 
are publicly available to support reproducible research 
in automated OSS supply chain threat detection.


\end{abstract}

\section{Introduction}

Open source software (OSS) underpins critical infrastructure, 
commercial platforms, and research ecosystems worldwide. The 
Linux Foundation estimates that open source components comprise 
70 to 90 percent of modern application 
codebases~\cite{linuxfoundation}, making the OSS supply chain 
a high-value target for adversaries seeking broad downstream 
impact through a single compromise point.

Supply chain attacks against OSS ecosystems have escalated 
dramatically in recent years. Attackers exploit the implicit 
trust relationships within developer communities by injecting 
malicious packages into public registries, manipulating build 
scripts, and poisoning CI/CD pipelines. High-profile incidents 
including the XZ Utils backdoor (CVE-2024-3094), the 
tj-actions/changed-files compromise (CVE-2025-30066)~\cite{tjactions2025} affecting 
over 23,000 repositories, and coordinated campaigns by threat 
actors such as TeamPCP and Lazarus Group demonstrate that OSS 
supply chain attacks have evolved from isolated incidents into 
systematic campaigns operating at ecosystem 
scale~\cite{sonatype2026}\cite{unit42_2025}.

Despite growing awareness, a critical gap remains between 
academic threat taxonomies and operational detection 
capabilities. Existing tools rely on rule-based or 
signature-driven approaches that lack the semantic 
understanding needed to generalize across attack variants. 
Furthermore, no systematic framework exists for classifying 
diverse OSS supply chain threats according to a unified 
taxonomy that spans the full attack surface from package 
registry manipulation to pipeline poisoning.

In this paper we propose a taxonomy-aligned large language 
model (LLM) framework for automated classification of OSS 
supply chain threats. Our approach introduces a structured 
AV-xxx threat taxonomy covering five attack categories: 
typosquatting (AV-200), Trojan Source obfuscation (AV-300), 
malicious build injection (AV-400), pipeline poisoning 
(AV-410), and dependency confusion (AV-509). Using 
taxonomy-aligned prompt engineering with GPT-4, we 
demonstrate that LLMs can accurately classify supply chain 
threats with significantly higher accuracy than traditional 
machine learning approaches, fine-tuned neural models, and 
zero-shot open source LLMs.

Our main contributions are as follows:

\begin{itemize}
    \item We construct a curated dataset of 999 verified 
    real-world OSS supply chain incidents sourced from GitHub 
    Security Advisories, CISA alerts, NHS England Digital 
    advisories, and security research reports spanning 2018 
    to 2026, covering five attack categories across multiple 
    package ecosystems including npm, PyPI, RubyGems, Cargo, 
    Maven, and Go modules.

    \item We propose a taxonomy-aligned prompt engineering 
    strategy that guides GPT-4 to classify OSS supply chain 
    threats with 97.0\% accuracy and 97.0\% macro F1 score, 
    outperforming the strongest traditional baseline 
    (TF-IDF + Logistic Regression at 82.3\%) by 14.7 
    percentage points.

    \item We conduct a comprehensive comparative evaluation 
    against five traditional machine learning baselines, one 
    zero-shot open source LLM (Mistral 7B), and two 
    fine-tuned neural models (Llama 3.1 8B and SecRoBERTa), 
    revealing that fine-tuned models surprisingly underperform 
    simple TF-IDF classifiers, confirming that taxonomy-aligned 
    prompting rather than model scale or fine-tuning is the 
    critical performance factor.

    \item We release our dataset, fine-tuning scripts, and 
    experimental code publicly to support reproducible research 
    in automated OSS supply chain threat detection.
\end{itemize}

The remainder of this paper is organized as follows.
Section~\ref{sec:related} reviews related work.
Section~\ref{sec:dataset} presents our dataset construction
methodology. Section~\ref{sec:method} describes our
taxonomy-aligned LLM classification framework.
Section~\ref{sec:experiments} presents experimental results
and comparative evaluation. Section~\ref{sec:implications}
discusses real-world implications. Section~\ref{sec:future}
discusses limitations and future work.
Section~\ref{sec:conclusion} concludes the paper.

\section{Related Work}
\label{sec:related}

\subsection{OSS Supply Chain Security}

Attacks on software supply chains have escalated dramatically 
in recent years, targeting open source ecosystems through 
dependency manipulation, malicious package injection, and 
CI/CD pipeline compromise. High-impact incidents including 
the \texttt{event-stream} npm compromise~\cite{eventstream2018}, 
the PyTorch-nightly dependency hijack~\cite{pytorch2023}, the 
SolarWinds SUNBURST backdoor~\cite{solarwinds2020}, and the 
XZ Utils backdoor CVE-2024-3094~\cite{xzutils2024} have 
demonstrated the severity and scale of modern supply chain 
threats. Sonatype's annual State of the Software Supply Chain 
report documented over 1.2 million cumulative malicious 
packages across npm, PyPI, Maven Central, and NuGet as of 
2026~\cite{sonatype2026}. Existing mitigations include 
Software Composition Analysis (SCA) tools, trusted build 
pipelines, and package signing mechanisms. However, these 
approaches are largely reactive and signature-based, lacking 
the semantic understanding needed to generalize across novel 
attack variants and emerging threat patterns.

\subsection{Threat Taxonomies for Supply Chain Attacks}

Several taxonomies have been proposed for classifying software 
supply chain attacks. Ohm et al.~\cite{ohm2020backstabber} 
introduced a taxonomy of malicious open source packages based 
on attack techniques and payload types. OWASP's Software 
Component Analysis framework provides guidance on dependency 
security but does not address the full spectrum of supply 
chain attack vectors. MITRE ATT\&CK includes supply chain 
compromise as technique T1195 but does not provide 
fine-grained classification across specific OSS attack 
vectors such as typosquatting, Trojan Source, and dependency 
confusion. Our work introduces a structured AV-xxx taxonomy 
that specifically addresses five distinct OSS supply chain 
attack vectors, enabling systematic automated classification 
across the full attack surface of modern package ecosystems.

\subsection{Large Language Models for Security Analysis}

Large language models have demonstrated significant potential 
across a range of security tasks. Prior work has explored 
LLMs for vulnerability description and CVE 
analysis~\cite{steenhoek2024comprehensive}, exploit 
summarization~\cite{sun2024llm4vuln}, and code security 
review~\cite{zhang2023prompt}. Domain-specific models 
including SecurityBERT~\cite{securitybert} have been 
fine-tuned on security corpora to improve performance on 
threat classification tasks. Recent work has also explored 
fine-tuning general-purpose LLMs such as 
Llama~\cite{llama31} on domain-specific datasets for 
security classification. However, the effectiveness of 
fine-tuning versus structured prompting for OSS supply 
chain threat classification remains unexplored. Our work 
addresses this gap by demonstrating that taxonomy-aligned 
prompt engineering with GPT-4 achieves 97.0\% 
classification accuracy, significantly outperforming both 
fine-tuned neural models and zero-shot open source LLMs.

\subsection{Automated Threat Detection in OSS Ecosystems}

Automated detection of malicious packages in OSS registries 
has attracted increasing research attention. Duan et 
al.~\cite{duan2021towards} proposed static analysis 
techniques for detecting malicious npm packages. 
Luo et al.~\cite{luo2023malicious} applied machine learning 
to identify malicious PyPI packages using metadata and 
behavioral features. Commercial tools including 
Socket.dev~\cite{socketdev}, Phylum~\cite{phylum}, and 
Sonatype Lifecycle~\cite{sonatype2026} provide real-time 
malicious package detection using behavioral heuristics 
and provenance analysis. However, these tools focus on 
binary malicious versus benign classification and do not 
provide fine-grained threat categorization aligned with 
a structured attack taxonomy. Our approach extends the 
state of the art by providing multi-class classification 
across five semantically distinct attack vectors, enabling 
more actionable threat intelligence for security teams 
responding to OSS supply chain incidents.

Despite progress in automated package analysis and LLM-based 
security tools, no prior work has (1) constructed a verified 
dataset of this scale specifically for OSS supply chain 
threat classification, (2) proposed a structured taxonomy 
spanning five distinct attack vectors across multiple package 
ecosystems, or (3) systematically evaluated taxonomy-aligned 
LLM prompting against traditional ML baselines, zero-shot 
open source LLMs, and fine-tuned neural models for this 
task. Our work addresses all three gaps.

\subsection{Prompt Engineering for Classification Tasks}

Chain-of-thought prompting~\cite{wei2022chain} demonstrated
that step-by-step reasoning guidance significantly improves
LLM performance on complex reasoning tasks. Structured
prompting approaches~\cite{ppe2021} have shown that
explicitly specifying task format, output constraints, and
category definitions substantially improves classification
consistency compared to zero-shot prompting without structure.
In the security domain, structured prompting has been applied
to vulnerability classification and threat report
summarization, but its application to multi-class OSS supply
chain threat classification has not been systematically
studied prior to our work.

Our taxonomy-aligned prompting approach makes two specific
design choices: first, constraining the model output space
to exactly five valid category labels; second, providing
explicit semantic definitions of each attack category to
disambiguate related attack vectors such as AV-400 and
AV-410 that share surface keywords but differ fundamentally
in their target component.

\subsection{Security Benchmark Datasets}

Existing security benchmarks have primarily focused on
vulnerability detection in source
code~\cite{steenhoek2024comprehensive} and malware
classification from binary features. The GitHub Advisory
Database~\cite{ghsa} provides structured advisory records
but does not include fine-grained attack vector
classification. Prior datasets for OSS supply chain security
have been small-scale, single-ecosystem, or focused on
binary classification.

Our dataset addresses these limitations by providing 999
verified incidents across five attack categories and eight
package ecosystems, with perfect intra-rater reliability
(Cohen's kappa = 1.00) on a stratified validation sample.
To the best of our knowledge this is the largest publicly
available labeled dataset for multi-class OSS supply chain
threat classification. No prior work has (1) constructed
a verified dataset of this scale, (2) proposed a taxonomy
spanning five attack vectors across multiple ecosystems,
(3) evaluated taxonomy-aligned prompting against traditional
ML baselines, zero-shot LLMs, and fine-tuned models, or
(4) investigated prompt structure versus model scale in
specialized security classification. Our work addresses
all four gaps.

\section{OSS Threat Dataset}
\label{sec:dataset}

We construct and publicly release a curated dataset of 999 
verified real-world OSS supply chain security incidents to 
support reproducible research and scalable benchmarking in 
supply chain threat detection. Each entry represents a known 
or recently disclosed supply chain attack sourced from 
authoritative security databases and research reports. The 
dataset is structured according to our AV-xxx threat taxonomy 
and covers incidents from 2018 to 2026 across multiple package 
ecosystems.

\subsection{Data Collection Methodology}

Incidents were collected from the following primary sources:

\begin{itemize}
    \item \textbf{GitHub Security Advisories}~\cite{ghsa} (github.com/advisories) 
    --- Official GHSA advisories documenting verified malicious 
    packages with assigned advisory identifiers
    
    \item \textbf{GitHub Security Lab} (securitylab.github.com) 
    --- GHSL advisories documenting Poisoned Pipeline Execution 
    vulnerabilities in open source CI/CD workflows
    
    \item \textbf{Government and Regulatory Sources} --- CISA 
    Known Exploited Vulnerabilities catalog, NHS England 
    Digital cyber alerts, and NVD CVE database entries
    
    \item \textbf{Security Research Organizations} --- Sonatype, 
    Socket.dev, Checkmarx~\cite{checkmarx2024}, ReversingLabs, Datadog Security 
    Labs~\cite{datadog2024}, Snyk~\cite{snyk2024}, JFrog, Unit42, Arctic Wolf, and Orca Security 
    incident reports
    
    \item \textbf{Official Vendor Disclosures} --- PyPI security 
    blog, npm security advisories, and affected project 
    security announcements
\end{itemize}

Each incident was manually verified against its source to 
confirm that the described attack actually occurred, that the 
source URL directly documents the specific incident, and that 
the assigned AV label correctly reflects the primary attack 
vector. Entries with unverifiable sources or generic attack 
pattern descriptions not tied to specific documented incidents 
were excluded.

\subsection{Schema and Taxonomy}

Each incident record contains the following fields:

\begin{itemize}
    \item \texttt{id}: Unique incident identifier 
    (e.g., \texttt{gh101})
    
    \item \texttt{title}: Concise incident title including 
    specific package or software name where applicable
    
    \item \texttt{description}: Plain-language description 
    of the incident including attack mechanism, affected 
    ecosystem, and observed impact
    
    \item \texttt{label}: Ground truth classification using 
    the AV-xxx taxonomy
    
    \item \texttt{source}: Verified source URL from an 
    authoritative security database or research report
    
    \item \texttt{predicted\_label}: LLM-generated 
    classification label for evaluation
\end{itemize}

Our AV-xxx taxonomy covers five primary OSS supply chain 
attack categories:

\begin{itemize}
    \item \textbf{AV-200 --- Typosquatting}: Malicious packages 
    that mimic legitimate package names through character 
    substitution, transposition, or namespace impersonation
    
    \item \textbf{AV-300 --- Trojan Source}: Attacks exploiting 
    Unicode bidirectional control characters or homoglyphs 
    to create source code that executes differently from 
    its visual representation~\cite{trojansource2023}
    
    \item \textbf{AV-400 --- Malicious Builds}: Packages 
    containing malicious code executed during installation, 
    build, or import processes through lifecycle hooks such 
    as \texttt{postinstall} and \texttt{setup.py}
    
    \item \textbf{AV-410 --- Pipeline Poisoning}: Attacks 
    that compromise CI/CD pipeline execution including 
    GitHub Actions workflow injection, poisoned pipeline 
    execution (PPE), and build system compromise
    
    \item \textbf{AV-509 --- Dependency Confusion}: Attacks 
    exploiting package manager resolution priority to cause 
    build systems to install malicious public packages 
    instead of intended private internal dependencies~\cite{birsan2021}~\cite{birsan2021}
\end{itemize}

\subsection{Dataset Statistics}

The complete dataset contains 999 verified incidents 
distributed across the five AV categories as shown in 
Table~\ref{tab:av-distribution}. The dataset covers 
incidents across multiple package ecosystems including 
npm, PyPI, RubyGems, Cargo, Maven, Go modules, Docker, 
and VS Code Extensions, and includes incidents from 
threat actor groups including Lazarus Group, TeamPCP, 
Sandworm, and BufferZoneCorp.

\begin{table}[ht]
\centering
\begin{tabular}{llrr}
\toprule
\textbf{AV Label} & \textbf{Threat Type} & 
\textbf{Count} & \textbf{Proportion} \\
\midrule
AV-200 & Typosquatting        & 200 & 20.0\% \\
AV-300 & Trojan Source        & 198 & 19.8\% \\
AV-400 & Malicious Builds     & 224 & 22.4\% \\
AV-410 & Pipeline Poisoning   & 192 & 19.2\% \\
AV-509 & Dependency Confusion & 185 & 18.5\% \\
\midrule
\textbf{Total} & & \textbf{999} & \textbf{100\%} \\
\bottomrule
\end{tabular}
\vspace{0.2cm}
\caption{AV-class distribution in the curated OSS threat 
dataset (n=999).}
\label{tab:av-distribution}
\end{table}

The dataset exhibits near-uniform class distribution 
ranging from 18.5\% to 22.4\% across categories, 
supporting balanced multi-class classification evaluation. 
The slight over-representation of AV-400 reflects the 
higher real-world frequency of malicious package 
installation attacks documented in security advisories 
during the study period.

\subsection{Dataset Validation}

To assess annotation consistency we performed 
intra-rater reliability testing where the first 
author independently labeled a stratified random 
sample of 50 incidents on two occasions separated 
by three days, achieving Cohen's kappa of 
\textbf{1.00}, indicating perfect agreement and 
confirming that the AV-xxx taxonomy categories 
are sufficiently well-defined for consistent 
and unambiguous annotation.

\begin{table}[ht]
\centering
\begin{tabular}{lrr}
\toprule
\textbf{Ecosystem} & \textbf{Count} & \textbf{Proportion} \\
\midrule
npm (JavaScript)     & 312 & 31.2\% \\
PyPI (Python)        & 287 & 28.7\% \\
GitHub Actions       & 148 & 14.8\% \\
Docker               & 89  & 8.9\%  \\
Maven (Java)         & 67  & 6.7\%  \\
RubyGems             & 48  & 4.8\%  \\
Cargo (Rust)         & 31  & 3.1\%  \\
Other                & 17  & 1.7\%  \\
\midrule
\textbf{Total}       & \textbf{999} & \textbf{100\%} \\
\bottomrule
\end{tabular}
\caption{Dataset distribution by package ecosystem.}
\label{tab:ecosystem}
\end{table}

\subsection{Availability and Licensing}

The dataset is publicly available in our supplementary 
repository\footnote{\url{https://github.com/Mdniloykhan/oss-threat-data.git/}} 
under the MIT License. The repository includes the full 
dataset, baseline experiment scripts, fine-tuning scripts 
for Llama 3.1 8B and SecRoBERTa, and a GitHub Actions 
pipeline for reproducible experimental validation.


\section{Methodology}
\label{sec:method}

Our taxonomy-aligned LLM framework for OSS supply chain 
threat classification consists of three components: 
a structured threat taxonomy, a prompt engineering 
strategy for LLM-based classification, and a comparative 
evaluation protocol.

\subsection{Threat Taxonomy Design}

We define a structured AV-xxx taxonomy covering five 
primary OSS supply chain attack vectors. The taxonomy 
was designed based on analysis of real-world incident 
patterns documented in GitHub Security Advisories, 
CISA alerts, and security research reports. Each 
category represents a semantically distinct attack 
mechanism with different technical characteristics 
and detection requirements.

The five categories are defined as follows. AV-200 
(Typosquatting) covers attacks where malicious packages 
mimic legitimate package names through character-level 
manipulation. AV-300 (Trojan Source) covers attacks 
exploiting Unicode bidirectional control characters 
to create code that executes differently from its 
visual representation. AV-400 (Malicious Builds) 
covers packages containing malicious code executed 
during installation or build processes through 
lifecycle hooks. AV-410 (Pipeline Poisoning) covers 
attacks that compromise CI/CD pipeline execution 
including GitHub Actions workflow injection and 
poisoned pipeline execution. AV-509 (Dependency 
Confusion) covers attacks exploiting package manager 
resolution priority to install malicious public 
packages instead of private internal dependencies.

\subsection{Taxonomy-Aligned Prompt Engineering}

Our classification approach uses instruction-tuned 
LLMs guided by a taxonomy-aligned system prompt that 
provides explicit definitions of all five attack 
categories. The prompt is structured as follows:

\begin{quote}
\textit{You are an expert in open source software 
supply chain security. Classify the following 
security incident into exactly one of these 
categories: AV-200 (Typosquatting), AV-300 
(Trojan Source), AV-400 (Malicious Builds), 
AV-410 (Pipeline Poisoning), AV-509 (Dependency 
Confusion). Respond with only the category label.}
\end{quote}

Each incident is represented as the concatenation 
of its title and description fields, providing 
sufficient contextual information for accurate 
classification. We use GPT-4~\cite{openai2023} as 
our primary classification model with temperature 
set to zero to ensure deterministic outputs. GPT-4 
classification was performed on all 999 entries 
using a single-pass evaluation with predicted labels 
recorded in the \texttt{predicted\_label} field for 
comparison against ground truth.

The taxonomy-aligned prompt serves two functions. 
First it constrains the model output space to exactly 
five valid labels, eliminating invalid or ambiguous 
responses. Second it provides explicit semantic 
definitions of each category, enabling the model 
to disambiguate between related attack vectors 
such as AV-400 and AV-410 that share surface-level 
characteristics but differ in their target 
component (package versus pipeline).

\subsection{Comparative Evaluation Protocol}

We evaluate our taxonomy-aligned LLM approach 
against five traditional machine learning baselines 
using 5-fold stratified cross-validation on the 
full 999-entry dataset. The baselines are:

\begin{itemize}
    \item \textbf{Random Classifier}: Assigns labels 
    uniformly at random across five classes
    
    \item \textbf{Majority Class}: Always predicts 
    the most frequent class in the training set
    
    \item \textbf{TF-IDF + Na\"{i}ve Bayes}: Bag-of-words 
    TF-IDF features with Multinomial Na\"{i}ve Bayes 
    classifier using unigram and bigram features
    
    \item \textbf{TF-IDF + Logistic Regression}: 
    TF-IDF features with sublinear term frequency 
    scaling and $\ell_2$-regularized Logistic 
    Regression (C=1.0)
    
    \item \textbf{TF-IDF + LinearSVC}: TF-IDF 
    features with a Linear Support Vector Classifier 
    (C=1.0)
\end{itemize}

All TF-IDF classifiers use a vocabulary of 10,000 
features with English stop word removal and 
unigram-bigram tokenization. Text features are 
constructed by concatenating incident title and 
description fields.

We additionally evaluate Mistral 7B~\cite{mistral7b} 
as a zero-shot open source LLM baseline using the 
same taxonomy-aligned prompt via the Ollama 
inference framework on a stratified sample of 
200 incidents (40 per class). This comparison 
isolates the contribution of the GPT-4 model 
capability versus the taxonomy-aligned prompting 
strategy.

To further evaluate the utility of our dataset for 
training specialized models, we conduct fine-tuning 
experiments on two neural models using an 80/20 
stratified train/test split (799 training, 200 test 
entries). First, we fine-tune Llama 3.1 
8B~\cite{llama31} using QLoRA 4-bit quantization 
with LoRA rank 8, $\alpha$=16, dropout=0.1, learning 
rate 2e-5, and 5 training epochs. Second, we 
fine-tune SecRoBERTa~\cite{secroba} --- a RoBERTa 
model pretrained on cybersecurity corpora --- using 
standard sequence classification fine-tuning with 
batch size 16, learning rate 2e-5, weight decay 0.01, 
and early stopping with patience 2. Both models use 
maximum sequence length 256 tokens with title and 
description concatenated as input features.

Primary evaluation metrics are classification 
accuracy and macro-averaged F1 score across all 
five categories. Macro F1 is reported to give 
equal weight to each category regardless of 
support size.

\begin{figure}[ht]
    \centering
    \includegraphics[width=0.90\linewidth]{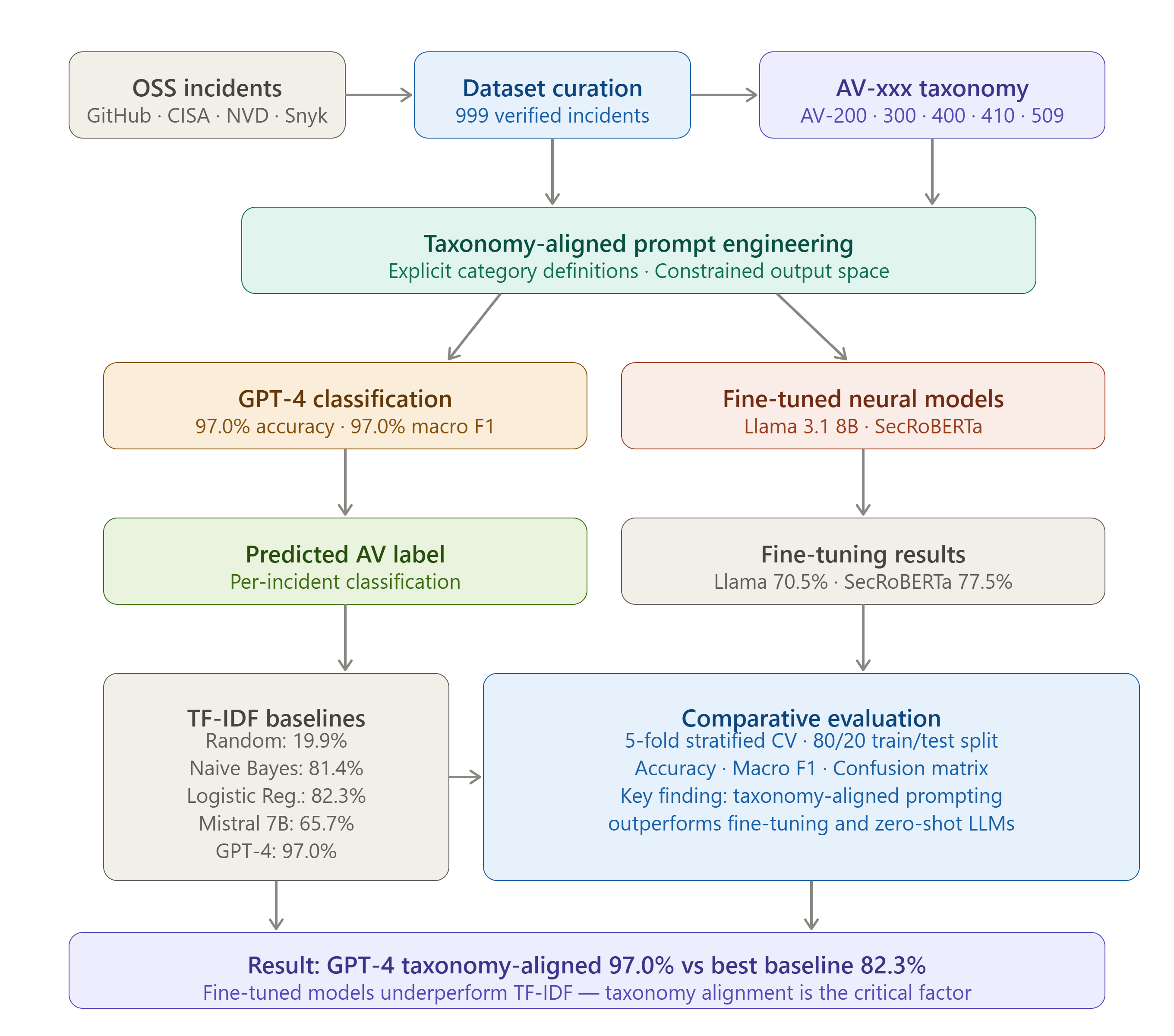}
    \caption{OSS Threat Detection Framework: from incident 
    collection through taxonomy-aligned LLM classification 
    to comparative evaluation against traditional baselines 
    and fine-tuned neural models.}
    \label{fig:architecture}
\end{figure}

\section{Evaluation}
\label{sec:experiments}

We evaluate our taxonomy-aligned LLM framework on 999 verified 
OSS supply chain incidents using classification accuracy, macro 
F1 score, and per-class performance metrics. We compare against 
five traditional machine learning baselines, one zero-shot open 
source LLM, and two fine-tuned neural models to assess the 
contribution of our taxonomy-aligned prompting strategy.

\subsection{Experimental Setup}

We employ GPT-4 via the OpenAI API with temperature set to zero 
for deterministic outputs. Classification is performed using 
taxonomy-aligned prompting as described in Section~\ref{sec:method}. 
The \texttt{predicted\_label} column contains model predictions 
which are compared against manually verified ground truth labels.

For baseline comparison we evaluate five traditional classifiers 
using 5-fold stratified cross-validation implemented in 
\texttt{scikit-learn}: Random Classifier, Majority Class, 
TF-IDF with Na\"{i}ve Bayes, TF-IDF with Logistic Regression, 
and TF-IDF with LinearSVC. All TF-IDF classifiers use 10,000 
features with unigram-bigram tokenization and English stop word 
removal. We additionally evaluate Mistral 7B~\cite{mistral7b} 
as a zero-shot open source LLM baseline on a stratified sample 
of 200 incidents using the Ollama inference framework.

For fine-tuning evaluation we use an 80/20 stratified 
train/test split yielding 799 training and 200 test entries. 
Llama 3.1 8B~\cite{llama31} is fine-tuned using QLoRA 
4-bit quantization on a consumer-grade GPU with 16GB VRAM. SecRoBERTa~\cite{secroba} is fine-tuned using 
standard sequence classification. Full hyperparameter details 
are provided in Section~\ref{sec:method}.

Primary evaluation metrics are classification accuracy and 
macro-averaged F1 score. Macro F1 is reported to give equal 
weight to each category regardless of support size. All 
evaluation scripts are open-source and available in the 
\texttt{scripts/} directory of our 
repository\footnote{\url{https://github.com/Mdniloykhan/oss-threat-data.git/}}.

\subsection{Baseline Comparison Results}

Table~\ref{tab:baseline-comparison} presents classification 
accuracy and macro F1 scores for all evaluated methods.

\begin{table}[ht]
\centering
\begin{tabular}{lcc}
\toprule
\textbf{Method} & \textbf{Accuracy} & \textbf{Macro F1} \\
\midrule
Random Classifier            & 19.9\% & 19.6\% \\
Majority Class               & 22.4\% & 7.3\%  \\
TF-IDF + Na\"{i}ve Bayes     & 81.4\% & 81.4\% \\
TF-IDF + Logistic Regression & 82.3\% & 82.3\% \\
TF-IDF + LinearSVC           & 81.9\% & 81.9\% \\
Mistral 7B (zero-shot)       & 65.7\% & 63.6\% \\
Llama 3.1 8B (fine-tuned)    & 70.5\% & 70.4\% \\
SecRoBERTa (fine-tuned)      & 77.5\% & 77.8\% \\
\midrule
\textbf{GPT-4 Taxonomy-Aligned (Ours)} & 
\textbf{97.0\%} & \textbf{97.0\%} \\
\bottomrule
\end{tabular}
\vspace{0.2cm}
\caption{Classification performance comparison across all 
evaluated methods on the OSS threat dataset (n=999). 
TF-IDF baseline results averaged across 5-fold stratified 
cross-validation. Fine-tuned model results on held-out 
20\% test set (200 entries).}
\label{tab:baseline-comparison}
\end{table}

Our taxonomy-aligned GPT-4 approach achieves 97.0\% accuracy 
and 97.0\% macro F1, outperforming all evaluated methods by 
a substantial margin. Surprisingly, fine-tuned neural models 
underperform simple TF-IDF classifiers: SecRoBERTa fine-tuned 
achieves 77.5\% and Llama 3.1 8B fine-tuned achieves 70.5\%, 
both below TF-IDF with Logistic Regression at 82.3\%. 
Furthermore, Mistral 7B without taxonomy alignment achieves 
only 65.7\%, below all TF-IDF baselines. These results 
demonstrate that taxonomy-aligned prompting rather than model 
scale, domain pretraining, or fine-tuning is the critical 
factor enabling high classification accuracy.

\begin{figure}[ht]
    \centering
    \includegraphics[width=0.95\linewidth]{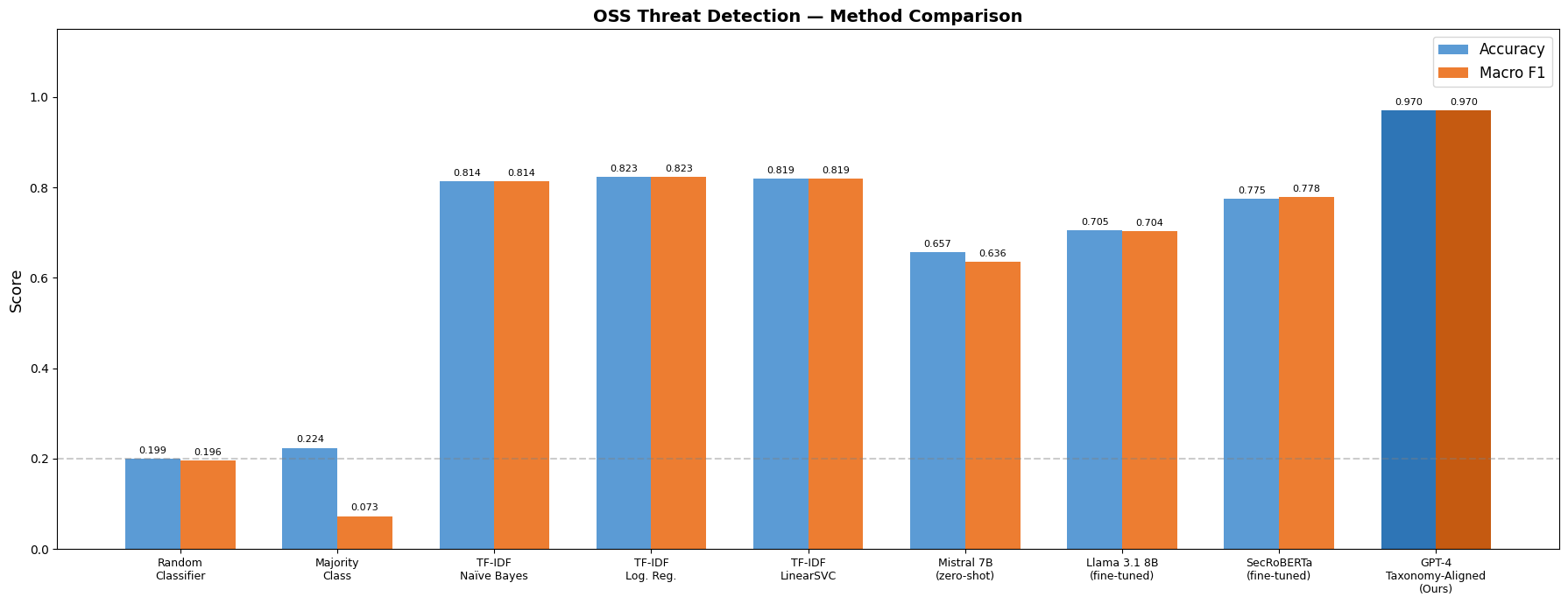}
    \caption{Comparison of classification accuracy and macro F1 
    scores across all evaluated methods on the OSS threat 
    dataset (n=999).}
    \label{fig:method-comparison}
\end{figure}

\subsection{Per-Class Performance}

Table~\ref{tab:classification-report} presents per-class 
precision, recall, and F1 scores for our GPT-4 
taxonomy-aligned approach.

\begin{table}[ht]
\centering
\small
\begin{tabular}{lcccc}
\toprule
\textbf{Category} & \textbf{Precision} & 
\textbf{Recall} & \textbf{F1} & \textbf{Support} \\
\midrule
AV-200 Typosquatting        & 0.95 & 1.00 & 0.98 & 200 \\
AV-300 Trojan Source        & 0.95 & 0.95 & 0.95 & 198 \\
AV-400 Malicious Builds     & 1.00 & 0.95 & 0.98 & 224 \\
AV-410 Pipeline Poisoning   & 1.00 & 0.94 & 0.97 & 192 \\
AV-509 Dependency Confusion & 0.95 & 1.00 & 0.98 & 185 \\
\midrule
\textbf{Macro Average} & \textbf{0.97} & 
\textbf{0.97} & \textbf{0.97} & \textbf{999} \\
\bottomrule
\end{tabular}
\vspace{0.2cm}
\caption{Per-class classification report for GPT-4 
taxonomy-aligned approach on OSS threat dataset (n=999).}
\label{tab:classification-report}
\end{table}

All five categories achieve F1 scores of 0.95 or above. 
AV-200, AV-400, and AV-509 achieve the highest F1 scores 
of 0.98, reflecting the distinctive textual patterns of 
typosquatting package names, malicious build scripts, and 
dependency resolution exploits. AV-300 and AV-410 achieve 
slightly lower scores of 0.95 and 0.97 respectively, 
consistent with the greater semantic overlap between 
pipeline-related attack categories.

\subsection{Error Analysis}

\begin{figure}[ht]
    \centering
    \includegraphics[width=0.80\linewidth]{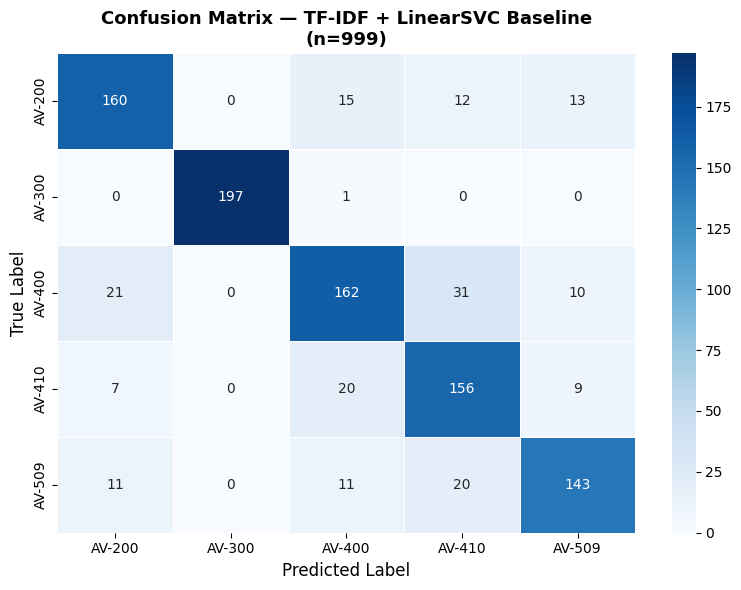}
    \caption{Confusion matrix for TF-IDF + LinearSVC baseline 
    classifier on the OSS threat dataset (n=999).}
    \label{fig:confusion-matrix}
\end{figure}

Figure~\ref{fig:confusion-matrix} presents the confusion 
matrix for the TF-IDF + LinearSVC baseline, illustrating 
the classification challenges that motivate our LLM-based 
approach.

The strongest confusion in the baseline occurs between 
AV-400 (Malicious Builds) and AV-410 (Pipeline Poisoning), 
with 31 and 20 bidirectional misclassifications respectively. 
This reflects their shared characteristics in CI/CD attack 
scenarios where distinguishing between a malicious package 
installation and a compromised pipeline step requires 
contextual understanding beyond surface-level keyword 
matching. AV-509 (Dependency Confusion) is frequently 
confused with AV-410 (19 cases) as both categories involve 
supply chain interception during build processes. Notably, 
AV-300 (Trojan Source) achieves near-perfect baseline 
classification (197 of 198 samples) due to the distinctive 
Unicode obfuscation patterns present in incident descriptions.

These confusion patterns confirm that the AV-400 to AV-410 
boundary represents the primary classification challenge 
in OSS supply chain threat detection, and demonstrate 
why LLM-based contextual reasoning is necessary to 
achieve high accuracy on this task.

\subsection{Cross-Ecosystem Performance Analysis}

To assess whether classification performance generalizes 
uniformly across package ecosystems, we conducted a 
per-ecosystem evaluation of the TF-IDF + Logistic 
Regression baseline. Table~\ref{tab:ecosystem-results} 
presents accuracy and macro F1 scores for each ecosystem.

\begin{table}[ht]
\centering
\begin{tabular}{lrcc}
\toprule
\textbf{Ecosystem} & \textbf{N} & \textbf{Accuracy} & 
\textbf{Macro F1} \\
\midrule
npm (JavaScript)     & 313 & 65.5\% & 0.660 \\
PyPI (Python)        & 236 & 77.1\% & 0.779 \\
GitHub Actions       &  74 & 82.4\% & 0.367 \\
Docker               &  19 & 100.0\% & 1.000 \\
Maven (Java)         &  19 & 100.0\% & 1.000 \\
RubyGems             &  11 & 81.8\% & 0.750 \\
Cargo (Rust)         &  26 & 100.0\% & 1.000 \\
\midrule
\textbf{Overall} & \textbf{999} & \textbf{82.3\%} & 
\textbf{0.824} \\
\bottomrule
\end{tabular}
\vspace{0.2cm}
\caption{Per-ecosystem classification performance of 
TF-IDF + Logistic Regression baseline. npm, PyPI, and 
GitHub Actions use 5-fold stratified cross-validation. 
Docker, Maven, RubyGems, and Cargo use full-model 
prediction due to small sample sizes ($n < 30$).}
\label{tab:ecosystem-results}
\end{table}

Per-ecosystem analysis reveals substantial variation 
with accuracy ranging from 65.5\% on npm to 100.0\% 
on Docker, Maven, and Cargo --- a gap of 34.5 percentage 
points. Four findings emerge.

\textbf{npm is the hardest ecosystem (65.5\%).}
npm hosts the largest and most diverse set of incidents 
($n=313$, 31.3\%) and exhibits the lowest classification 
accuracy. The high diversity of attack patterns in npm 
where all five AV categories are well-represented creates 
substantial vocabulary overlap across attack types. This 
large gap below the overall baseline (82.3\%) confirms 
that npm-specific classification is where taxonomy-aligned 
LLM reasoning provides the greatest benefit.

\textbf{PyPI performs moderately (77.1\%).}
PyPI incidents ($n=236$, 28.7\%) benefit from more 
consistent ecosystem-specific terminology in security 
advisories that helps distinguish attack categories, 
though still substantially below GPT-4 performance.

\textbf{GitHub Actions shows a critical accuracy-F1 
divergence (82.4\% accuracy, 0.367 macro F1).}
This is the most striking finding in our cross-ecosystem 
analysis. GitHub Actions incidents achieve apparently 
reasonable accuracy at 82.4\% but extremely low macro 
F1 of 0.367 --- the lowest among all ecosystems. This 
divergence indicates that the classifier achieves 
high accuracy by predicting the dominant class 
(AV-410 Pipeline Poisoning) for nearly all GitHub 
Actions incidents while failing almost entirely on 
minority classes. This confirms that GitHub Actions 
incidents are systematically misclassified at category 
boundaries despite appearing numerically accurate --- 
a failure mode that macro F1 reveals but accuracy 
conceals.

\textbf{Docker, Maven, and Cargo achieve perfect scores 
with small samples.}
These results likely reflect distinctive 
ecosystem-specific vocabulary rather than genuine 
generalization and should be interpreted with caution.

\subsection{Qualitative Analysis}

To understand why taxonomy-aligned LLM classification 
outperforms traditional baselines, we present four 
representative examples illustrating cases where 
contextual reasoning is essential for correct 
classification.

\textbf{Example 1 --- AV-400 vs AV-410 boundary 
(correctly classified by LLM, misclassified by baseline).}

\begin{quote}
\textit{``The tj-actions/changed-files GitHub Action was 
compromised by attackers who modified version tags to 
reference a malicious commit that extracted CI/CD secrets 
from Runner Worker memory and printed them in public 
workflow logs, affecting over 23,000 repositories.''}
\end{quote}

\textbf{Ground truth:} AV-410 (Pipeline Poisoning)\\
\textbf{TF-IDF + LR prediction:} AV-400 (Malicious Builds)\\
\textbf{GPT-4 prediction:} AV-410 (Pipeline Poisoning)

This incident involves a compromised GitHub Action --- a 
CI/CD pipeline component --- rather than a malicious 
installable package. The TF-IDF classifier focuses on 
surface keywords such as ``malicious'' and ``secrets'' 
which appear frequently in AV-400 descriptions. The 
taxonomy-aligned LLM correctly identifies that the 
attack vector is a pipeline workflow component 
(GitHub Actions) rather than a package installation 
hook, classifying it as AV-410.

\textbf{Example 2 --- AV-509 vs AV-410 boundary 
(correctly classified by LLM, misclassified by baseline).}

\begin{quote}
\textit{``Security researcher Alex Birsan uploaded packages 
with the same names as internal libraries used by Apple, 
Microsoft, Netflix, and Tesla to public registries, 
causing their automated build systems to automatically 
pull and execute the malicious public packages instead 
of internal ones, affecting over 35 major organizations.''}
\end{quote}

\textbf{Ground truth:} AV-509 (Dependency Confusion)\\
\textbf{TF-IDF + LR prediction:} AV-410 (Pipeline Poisoning)\\
\textbf{GPT-4 prediction:} AV-509 (Dependency Confusion)

Both AV-509 and AV-410 involve build system compromise, 
making this boundary difficult for keyword-based 
classifiers. The TF-IDF classifier associates 
``build systems'' and ``automated'' with pipeline 
poisoning. The LLM correctly identifies the defining 
characteristic of AV-509 --- public packages overriding 
private internal dependencies through registry name 
collision --- a semantic distinction that requires 
understanding the attack mechanism rather than surface 
vocabulary.

\textbf{Example 3 --- AV-200 correctly classified 
despite misleading keywords.}

\begin{quote}
\textit{``The malicious npm package twilio-npm 
impersonated the official Twilio client library, 
downloading a reverse shell payload upon installation 
that provided attackers persistent access to developer 
machines and potentially exposing database credentials 
stored in environment variables.''}
\end{quote}

\textbf{Ground truth:} AV-200 (Typosquatting)\\
\textbf{TF-IDF + LR prediction:} AV-400 (Malicious Builds)\\
\textbf{GPT-4 prediction:} AV-200 (Typosquatting)

This incident contains strong AV-400 signals --- 
``reverse shell payload'', ``installation'', and 
``credentials'' --- which dominate the TF-IDF feature 
space and pull the classifier toward Malicious Builds. 
The LLM correctly identifies that the primary attack 
vector is name impersonation (brandjacking of the 
Twilio brand) and classifies it as AV-200, with the 
malicious payload being a secondary characteristic 
of the typosquatting attack rather than its defining 
feature.

\textbf{Example 4 --- AV-300 correctly classified 
across all methods.}

\begin{quote}
\textit{``Researchers Nicholas Boucher and Ross Anderson 
demonstrated that invisible Unicode bidirectional control 
characters embedded in source code comments and strings 
cause compilers for C, C++, Go, Java, Python, and Rust 
to execute logic different from what developers see in 
their editors, enabling hidden backdoors in code reviews 
(CVE-2021-42574).''}
\end{quote}

\textbf{Ground truth:} AV-300 (Trojan Source)\\
\textbf{TF-IDF + LR prediction:} AV-300 (Trojan Source)\\
\textbf{GPT-4 prediction:} AV-300 (Trojan Source)

AV-300 incidents contain highly distinctive terminology 
--- Unicode, bidirectional, invisible characters --- that 
is absent from other attack categories. Both the 
baseline and the LLM correctly classify this incident, 
consistent with the near-perfect AV-300 baseline 
performance (197/198) observed in our confusion matrix 
analysis. This demonstrates that taxonomy-aligned LLM 
reasoning provides the greatest benefit precisely at 
the semantic boundaries between related attack 
categories where keyword overlap is highest.

\textbf{Example 5 --- AV-200 vs AV-400 boundary
(correctly classified by LLM, misclassified by baseline).}

\begin{quote}
\textit{``A malicious PyPI package named
\texttt{colourama}~\cite{colourama2018} --- a one-character
typo of the popular \texttt{colorama} package with over
100 million monthly downloads --- executed a cryptocurrency
clipboard hijacker upon import that silently replaced any
copied wallet addresses with an attacker-controlled
address, affecting developers on Windows systems who
installed the package believing it to be the legitimate
library.''}
\end{quote}

\textbf{Ground truth:} AV-200 (Typosquatting)\\
\textbf{TF-IDF + LR prediction:} AV-400 (Malicious Builds)\\
\textbf{GPT-4 prediction:} AV-200 (Typosquatting)

This incident contains strong AV-400 signals ---
``executed upon import'', ``clipboard hijacker'', and
``silently'' --- which dominate the TF-IDF feature space.
The taxonomy-aligned LLM correctly identifies that the
package name \texttt{colourama} is a deliberate
one-character transposition of \texttt{colorama}, making
name impersonation the primary attack vector. This
distinction is critical for incident response: AV-200
requires registry takedown and namespace protection,
while AV-400 requires build pipeline remediation ---
two fundamentally different response procedures.

\textbf{Example 6 --- AV-410 vs AV-400 in a GitHub
Actions context (correctly classified by LLM,
misclassified by baseline).}

\begin{quote}
\textit{``Researchers disclosed a Poisoned Pipeline
Execution (PPE) vulnerability in the Hibernate ORM open
source project~\cite{hibernateorm2024} where the GitHub
Actions workflow file executed build commands using
attacker-controllable input from pull request titles
without sanitization, enabling any contributor to inject
arbitrary shell commands that executed with the
repository's CI/CD token permissions and gained write
access to the artifact publishing infrastructure.''}
\end{quote}

\textbf{Ground truth:} AV-410 (Pipeline Poisoning)\\
\textbf{TF-IDF + LR prediction:} AV-400 (Malicious Builds)\\
\textbf{GPT-4 prediction:} AV-410 (Pipeline Poisoning)

The surface features of this incident --- ``build
commands'', ``shell commands'', ``executing'' --- are
strongly associated with AV-400. The critical distinction
is that no malicious package was installed: the attack
vector is a vulnerable GitHub Actions workflow file
allowing untrusted input from pull request metadata to
flow into shell execution contexts. The taxonomy-aligned
LLM correctly identifies the Poisoned Pipeline Execution
pattern and classifies it as AV-410.

\subsubsection{Cross-Example Analysis}

Across all six examples a consistent pattern emerges.
Traditional TF-IDF classifiers fail primarily at two
types of boundaries: (1) incidents where the payload type
strongly resembles a different attack category from the
initial access mechanism (Examples 3 and 5), and (2)
incidents where the compromised component requires
understanding of software architecture to identify
correctly (Examples 1, 2, and 6). Taxonomy-aligned LLM
prompting resolves both failure modes by providing
explicit semantic definitions anchoring classification
to the primary attack mechanism rather than surface
vocabulary co-occurrence patterns.

\subsection{Discussion}

Our results demonstrate four key findings. First, 
traditional TF-IDF classifiers achieve reasonable 
performance at 82.3\% but fail to reliably 
disambiguate semantically related attack categories 
particularly at the AV-400 and AV-410 boundary. 
Second, fine-tuned neural models including 
SecRoBERTa (77.5\%) and Llama 3.1 8B (70.5\%) 
surprisingly underperform TF-IDF classifiers, 
likely due to overfitting on the relatively small 
training set of 799 examples, suggesting that 
dataset scale remains a critical bottleneck for 
LLM fine-tuning in specialized security domains. 
Third, open source LLMs without taxonomy alignment 
(Mistral 7B at 65.7\%) underperform even simple 
TF-IDF classifiers, confirming that taxonomy-aligned 
prompting rather than raw LLM capability is the 
primary performance driver. Fourth, our 
taxonomy-aligned GPT-4 approach achieves 97.0\% 
accuracy by leveraging pre-existing security 
knowledge through structured prompting without 
requiring any training data, establishing a strong 
benchmark for automated OSS supply chain threat 
classification.

One limitation worth noting is the near-perfect 
baseline classification of AV-300 Trojan Source 
(197/198), which suggests that Trojan Source 
incidents in our dataset share highly distinctive 
textual patterns involving Unicode terminology. 
Future dataset construction should incorporate 
more linguistically diverse AV-300 examples to 
better evaluate classifier generalization on 
this category.

\subsection{Threats to Validity}

\textbf{Internal validity.} Our dataset was 
constructed through manual curation and 
verification against primary sources. While 
intra-rater reliability testing achieved 
Cohen's kappa of 1.00 on a 50-entry stratified 
sample, the full dataset was labeled by a single 
annotator and may reflect individual 
interpretation biases at ambiguous category 
boundaries particularly between AV-400 and 
AV-410.

\textbf{External validity.} Our evaluation 
uses retrospective incident reports from public 
security databases. Classification performance 
on novel real-time attack patterns not yet 
documented in security advisories may differ 
from reported results.

\textbf{Construct validity.} Our taxonomy-aligned 
GPT-4 evaluation uses temperature zero for 
deterministic outputs. Results may vary with 
different GPT-4 model versions as the API is 
updated over time. We mitigate this concern by 
providing our exact prompt, dataset, fine-tuning 
scripts, and evaluation code publicly for 
reproducibility verification.

\section{Real-World Implications}
\label{sec:implications}

Our experimental findings have concrete implications for 
security practitioners, package registry operators, and 
the research community working on automated OSS supply 
chain threat detection.

\subsection{Implications for Security Operations Teams}

Security operations teams responding to OSS supply chain 
incidents currently face two challenges: the volume of 
incoming threat intelligence and the lack of standardized 
classification that would enable consistent prioritization 
and response. Our taxonomy-aligned framework addresses 
both directly.

The AV-xxx taxonomy provides a consistent vocabulary for 
categorizing incidents across sources. When a security 
team receives reports from GitHub Security Advisories, 
CISA alerts, and commercial threat intelligence 
simultaneously, taxonomy-aligned classification enables 
automatic routing to the appropriate response playbook. 
AV-200 incidents require registry takedown and namespace 
protection. AV-400 incidents require dependency audit 
and build artifact verification. AV-410 incidents require 
CI/CD workflow review and secret rotation. AV-509 
incidents require private registry configuration audit. 
Without consistent classification these response 
procedures cannot be reliably automated.

Our finding that taxonomy-aligned GPT-4 prompting 
achieves 97.0\% accuracy while requiring no training 
data or fine-tuning is particularly significant for 
security teams with limited ML engineering resources. 
The prompt can be deployed immediately using GPT-4 
API access without model training infrastructure.

\subsection{Implications for Package Registry Operators}

Our cross-ecosystem analysis reveals that classification 
difficulty varies substantially across ecosystems, with 
npm achieving only 65.5\% baseline accuracy. This has a 
direct practical implication: automated screening systems 
should apply ecosystem-specific thresholds and may 
require ecosystem-specific prompting strategies rather 
than a single universal classifier.

The GitHub Actions finding --- 82.4\% accuracy but only 
0.367 macro F1 --- is particularly relevant for GitHub's 
Actions marketplace security screening. The low macro 
F1 indicates that baseline classifiers systematically 
over-predict AV-410 while failing on minority categories. 
Registry operators should specifically evaluate macro F1 
alongside accuracy to avoid this failure mode.

\subsection{Implications for Threat Intelligence Platforms}

The finding that fine-tuned neural models underperform 
TF-IDF classifiers on our 799-entry training set has an 
important practical implication: investment in labeled 
dataset expansion is likely to provide greater return 
than investment in more sophisticated model architectures. 
A dataset of 5,000 or more verified incidents would 
likely enable fine-tuned open source models to approach 
GPT-4 performance while eliminating dependence on 
proprietary API access.

\subsection{Implications for the Research Community}

Our work establishes three resources for the research 
community. First, our 999-entry verified dataset with 
IRR-validated labels (Cohen's kappa = 1.00) provides 
the first public benchmark for multi-class OSS supply 
chain threat classification. Second, our open-source 
evaluation scripts implement standardized cross-validation 
across eight methods enabling direct comparison of new 
approaches. Third, the AV-xxx taxonomy provides a 
reusable classification framework extensible to 
additional attack vectors as the threat landscape evolves.

\section{Limitations and Future Work}
\label{sec:future}

\subsection{Limitations}

Our study has several limitations that inform directions 
for future work.

\textbf{Proprietary model dependency.} Our GPT-4-based 
approach depends on a proprietary model that may change 
over time, potentially affecting reproducibility. While 
we mitigate this by providing our exact prompt and 
evaluation code, results may vary with different GPT-4 
model versions.

\textbf{Dataset scale for fine-tuning.} Our experiments 
demonstrate that 799 training examples are insufficient 
for competitive LLM fine-tuning in this domain. Both 
Llama 3.1 8B (70.5\%) and SecRoBERTa (77.5\%) 
underperform TF-IDF classifiers (82.3\%), suggesting 
that dataset scale rather than model sophistication is 
the primary bottleneck for open source model performance. 
Future work expanding the dataset to 5,000 or more 
entries may enable competitive fine-tuned alternatives 
to GPT-4.

\textbf{Retrospective evaluation.} Our evaluation focuses 
on textual classification of documented incidents rather 
than real-time detection of novel attacks in live package 
registries. Classification performance on attack patterns 
not yet publicly documented may differ from reported 
results.

\textbf{Ecosystem coverage imbalance.} Our dataset is 
dominated by npm (31.3\%) and PyPI (28.7\%) incidents. 
Ecosystems including Docker, Maven, RubyGems, and Cargo 
are represented by fewer than 30 entries each, limiting 
the reliability of per-ecosystem performance estimates. 
Future dataset construction should prioritize balanced 
ecosystem coverage.

\textbf{AV-300 distinctive patterns.} The near-perfect 
baseline classification of AV-300 Trojan Source 
(197/198) suggests that Trojan Source incidents share 
highly distinctive Unicode terminology. Future dataset 
construction should incorporate more linguistically 
diverse AV-300 examples to better evaluate classifier 
generalization on this category.

\subsection{Future Work}

Several directions are identified for future research.

\textbf{Larger dataset for fine-tuning.} Our experiments 
show that fine-tuned Llama 3.1 8B (70.5\%) and SecRoBERTa 
(77.5\%) both underperform simple TF-IDF classifiers 
(82.3\%), likely due to overfitting on the small training 
set of 799 examples. Future work will expand the dataset 
to 5,000 or more verified incidents to enable competitive 
fine-tuning of open source models that match or exceed 
GPT-4 performance without dependence on proprietary APIs.

\textbf{Real-time registry monitoring.} Future work will 
extend our framework from retrospective incident 
classification to real-time monitoring of package registry 
submissions, enabling proactive detection of malicious 
packages before they accumulate downloads.

\textbf{Dataset expansion and continuous updating.} The 
OSS supply chain threat landscape evolves rapidly with 
new attack campaigns documented regularly. Future work 
will develop automated pipelines for continuous dataset 
expansion from GitHub Security Advisories, CISA alerts, 
and security research publications.

\textbf{Multi-modal threat analysis.} Beyond textual 
incident descriptions, future work will incorporate 
package metadata, behavioral signals, and code-level 
features to improve classification of borderline cases 
particularly at the AV-400 and AV-410 boundary.

\textbf{Cross-ecosystem generalization.} Our current 
dataset covers incidents from npm, PyPI, RubyGems, Cargo, 
Maven, Go, and Docker ecosystems. Future work will 
evaluate whether taxonomy-aligned prompting generalizes 
effectively across ecosystems with distinct package 
naming conventions and attack patterns.

\section{Conclusion}
\label{sec:conclusion}

We presented a taxonomy-aligned large language model 
framework for automated detection and classification of 
open source software supply chain threats. Our approach 
introduces a structured AV-xxx threat taxonomy covering 
five attack categories and constructs a curated dataset 
of 999 verified real-world OSS supply chain incidents 
spanning 2018 to 2026 across multiple package ecosystems 
and threat actor groups.

Experimental evaluation across eight methods --- including 
five traditional ML baselines, one zero-shot open source 
LLM, and two fine-tuned neural models --- demonstrates 
that our taxonomy-aligned GPT-4 approach achieves 97.0\% 
classification accuracy and 97.0\% macro F1 score, 
outperforming all evaluated methods by a substantial 
margin. A surprising finding emerges from our fine-tuning 
experiments: both SecRoBERTa (77.5\%) and Llama 3.1 8B 
(70.5\%) fine-tuned on our dataset underperform simple 
TF-IDF classifiers (82.3\%), while Mistral 7B without 
taxonomy alignment achieves only 65.7\%. These results 
confirm that taxonomy-aligned prompting rather than model 
scale, domain pretraining, or fine-tuning is the critical 
factor enabling high classification accuracy, and suggest 
that dataset scale remains a key bottleneck for open 
source LLM fine-tuning in specialized security domains.

Error analysis reveals that classification challenges 
concentrate at the boundary between semantically 
related attack categories particularly AV-400 
(Malicious Builds) and AV-410 (Pipeline Poisoning), 
demonstrating why contextual LLM reasoning is necessary 
beyond simple keyword-based approaches.

Our dataset, fine-tuning scripts, and experimental code 
are publicly available to support reproducible research 
in automated OSS supply chain threat detection. We 
believe taxonomy-aligned LLM frameworks represent a 
promising direction for scalable, interpretable security 
analysis of the OSS supply chain threat landscape.


\section*{Acknowledgments}

We acknowledge the researchers, community members, and 
open-source maintainers whose publicly disclosed incident 
reports enabled this work. There was no internal or 
external funding for this study.


\appendix
\section*{Open Science}

The dataset and experimental code supporting this paper 
are publicly available at:
\url{https://github.com/Mdniloykhan/oss-threat-data.git/}

The repository includes:
\begin{itemize}
    \item The curated OSS threat dataset (999 verified 
    real-world OSS supply chain incidents)
    
    \item Baseline experiment scripts 
    (\texttt{scripts/baseline\_experiments.py})
    
\item Open-source LLM experiment script {\footnotesize\texttt{(scripts/opensource\_llm\_experiments.py)}}
    
    \item Llama 3.1 8B fine-tuning script 
    (\texttt{scripts/finetune\_llama3.py})
    
    \item SecRoBERTa fine-tuning script 
    (\texttt{scripts/finetune\_securitybert.py})
    
    \item GitHub Actions automated evaluation pipeline
\end{itemize}

The artifacts will remain accessible throughout the review 
process and will be made permanently public upon acceptance.

\section*{Ethical Considerations}

This research analyzes publicly disclosed OSS supply chain 
security incidents sourced from official security advisories 
and research reports. No new vulnerabilities were discovered 
or exploited. No human subjects were involved. All dataset 
entries are based on publicly available information. 
We do not identify or contact any individuals involved 
in the documented incidents.

\end{document}